\documentclass[aps,pra,preprint,showpacs,groupedaddress,endfloats]{revtex4}
\usepackage{graphicx}
\usepackage{setspace}

\newcommand{\be}{\begin{equation}}
\newcommand{\ee}{\end{equation}}
\newcommand{\bd}{\begin{displaymath}}
\newcommand{\ed}{\end{displaymath}}
\newcommand{\ba}{\begin{array}}
\newcommand{\ea}{\end{array}}
\newcommand{\bq}{\begin{eqnarray}}
\newcommand{\eq}{\end{eqnarray}}

\usepackage{doublespace}
\begin{document}

\title{Electron-molecule scattering calculations in a 3D finite element $R$-matrix approach}
\author{Stefano Tonzani}
\affiliation{JILA,  University of Colorado, Boulder, Colorado 80309-0440 }
\author{Chris H. Greene}
\affiliation{Department of Physics and JILA, University of Colorado, Boulder, Colorado 80309-0440}

\date{\today}

\begin{abstract}
We have implemented a three-dimensional finite element approach, based on tricubic
polynomials
in spherical coordinates, which solves the Schr\"{o}dinger equation for
scattering
of a low energy electron from a molecule, approximating the electron exchange as a local potential.
The potential is treated as a sum of three terms: electrostatic, exchange and
polarization.
The electrostatic term can be  extracted directly from $ab$ $initio$ codes
({\sc{GAUSSIAN 98}} in the work described here),
while the exchange term is approximated using different local density
functionals. A local polarization potential approximately describes the
long range attraction to the molecular target induced by the scattering
electron. 

\end{abstract}

\maketitle

 \section{ Introduction}
  Electron-molecule processes are important in many different areas of physics
and chemistry, for instance in cold plasmas (that are present in interstellar media
\cite{McCall:H3} and the
high atmospheric layers). They are also relevant in radiation damage to living tissue,
\cite{Sanche:DNA} and in surface physics and chemistry for example in
electron-beam induced chemistry. \cite{Ilemberger:PRL03}
Theoretical studies of electron collisions with molecular
targets have been carried out since the late 1970s (see for example
Refs. \onlinecite{{Morr_Coll:PRA78},
{Dill:2_80}}, while Ref. \onlinecite{Lane:rev80} presents an extensive review of the
state of the field up to  the early 1980s). 
Some of the adopted techniques include the Kohn variational principle,
\cite{Schneider:PRA88}
the Schwinger variational principle \cite{McKoy:PRA80} and the $R$-matrix method
\cite{Tennyson:H2} used in this study.  These methods have proven capable of
describing scattering from increasingly complex molecular targets.
\cite{{Gianturco:SF6},{Gianturco:c60}}

The need for a simple but general method to deal with electron
scattering by a 
polyatomic target, that does not utilize single center expansions or Gaussian
basis functions has led us to develop a new approach.
Much of our motivation derives from our goal of describing dissociative
recombination reactions and the role of Rydberg states in these
processes. 
Each of the techniques that constitute our
method  has been widely used  in the past, including the use of  finite elements
in scattering
processes, {\cite{{Shertz:PRA},{Wea:97}}} and the introduction of model
potentials to describe electron scattering. 
 \cite{{Morr_Coll:PRA78},{Dill:2_80}}
Nevertheless, to our knowledge, this is the first attempt to combine a three dimensional
finite element calculation with the $R$-matrix method. 
We hope that this approach can be used to
calculate quantum defect parameters, which can in turn describe vibrational-electronic coupling
in polyatomic molecules through an implementation of quantum defect
theory (QDT)
techniques.\cite{{Slava:03},{Jungen:PRL93},{Guberman:JCP91},{Takagi:DR}}

For this pilot study we describe the electron-molecule interaction through an
independent particle picture.
There are three main sources of interaction between a low-energy electron  and
a closed shell molecule: the direct electrostatic interaction, which is always the largest
contribution to the potential, the exchange interaction, which makes the
potential nonlocal and derives from  the antisymmetrization of the
wavefunction,
and a correlation and polarization term that describes the response of the
target to the continuum electron. 
The polarization term  is dealt with using a
simple long range polarization potential.

The exchange term, due
to its nonlocality, is the most complicated to model. We
reduce it to a local potential by adopting the widely employed local density
approximation (LDA). While this is a rather
crude approximation to this term in the potential, it is well-known that it gives surprisingly
realistic results; moreover it enables us to reduce the solution of the complicated scattering of an
electron from a multielectronic target to the solution of an effective one-body
Schr\"{o}dinger equation with a local potential. This description is expected
to be realistic only for closed-shell molecular targets.


Finite element techniques are well established as  flexible tools to solve
partial differential equations in different fields of physics, 
and in engineering. \cite{Bathe:book}

Their introduction to quantum mechanical calculations dates back to the
work of Shertzer and Botero, \cite{Shertz:PRA} who solved the
scattering equations as a few-body problem. They have also been implemented in
a study of two-electron photoejection from atoms. \cite{Meyer_Greene:PRL97} A  study that is closer in
spirit to ours is the one by Weatherford, $et.\; al.$ \cite{Wea:97}
which treats a simplified model Hamiltonian in a system possessing 
cylindrical symmetry, reducing the calculation to just two dimensions. 
The paper of Huo $et.\; al.$ \cite{Huo:PRA} uses instead an exact
representation of the exchange potential, using a Gaussian basis set at short
distance, and adopts finite elements only for the radial coordinate. 

\section{Theory}

\subsection{Electron scattering equations}
The electron molecule scattering problem, begins with
the full Hamiltonian of the system:
\be
\hat{H}=-\frac{1}{2}\sum_{i}\nabla^{2}_{r_i}-\frac{1}{2}\sum_{\alpha}\nabla^{2}_{R_{\alpha}}
-\sum_{i,\alpha}\frac{Z_{\alpha}}{\mid{\vec{r}_i-\vec{R}_{\alpha}}\mid}
+\sum_{\alpha>\beta}\frac{Z_{\alpha}
Z_{\beta}}{\mid{\vec{R}_\alpha-\vec{R}_{\beta}}\mid}
+\sum_{j>l}\frac{1}{\mid{\vec{r}_j-\vec{r}_l}\mid}.
\ee
This operator contains both the nuclear and electronic degrees of freedom,
indicated respectively with Greek and Latin indices. We
treat here the electronic problem alone, within the Born-Oppenheimer
approximation, namely freezing the nuclei in some definite configuration
(usually the equilibrium configuration) while solving for the electronic
wavefunction. The treatment of vibrations can be carried out
by repeating the electronic calculations for different values of
the nuclear positions, followed by vibrational averaging or a vibrational frame
transformation description. \cite{{Greene:85},{Fano:JOS75}}
It is now possible to write a wavefunction that depends
parametrically on the nuclear coordinates as an antisymmetrized product of the
target and scattering electron wavefunctions:
\be
\label{CI_wavefun:eq}
\Psi_{\gamma}={\cal{A}}\sum_{\gamma'}
\Phi_{\gamma'}(\bar{i},R) \phi_{0,\gamma'}(r_i)
\ee
where $\gamma$ represents the set of quantum numbers that fully describe the state of
the system, and the sum over $\gamma'$ allows for different configurations of
the compound system (target + scattered electron) to contribute. In Eq.
\ref{CI_wavefun:eq} $\bar{i}$ represents the coordinates of all the electrons
except the $i$-th.
 
If only the ground state configuration $\gamma '$ in this sum is retained, 
the approximation made is called static exchange. It is
possible to show \cite{Slater:book} in this case that the
($N+1$)-particle
Schr\"{o}dinger equation can be reduced to  $N+1$ single particle equations for
the individual orbitals. We are
interested in the orbital $\phi_0$ for 
the scattered electron, which obeys
\be
\label{1-body:eqn}
(-\nabla^{2}+V_{s}-E)\phi_{0}(\vec{r})= \sum_{j=1}^{N}
\phi_{j}(\vec{r})\int{d\vec{r'}\frac{\phi^{*}_{j}(\vec{r'})
\phi_{0}(\vec{r'})}{\mid\vec{r}
-\vec{r'}\mid}}
\ee
where the $\phi_{j}$ ($j \geq 1$) are the
target molecular orbitals. The electrostatic potential $V_{s}$ is the averaged Coulomb interaction of
the scattered electron with all the other electrons and the nuclei
\be
V_{s}(\vec{r})=\sum_{j=1}^{N}\int{d \vec{r'}
\frac{\phi_{j}^{*}(\vec{r'})\phi_{j}(\vec{r'})}{\mid\vec{r}-\vec{r'}\mid} } -
\sum_{\alpha}\frac{Z_{\alpha}}{\mid \vec{r}- \vec{R}_{\alpha}\mid}.
\ee
The term on the right hand side of Eq. \ref{1-body:eqn} is referred to as exchange potential.



\subsection{$R$-matrix method}
The $R$-matrix method is a well-established tool for 
problems where the continuum portion of the spectrum of a Hamiltonian must 
be treated. In its usual implementation, it involves diagonalization
of the (Bloch-modified) Hamiltonian operator in a box
subject to some fixed  boundary condition obeyed by the basis orbitals. The $R$-matrix box partitions
the space in two, with  an internal reaction zone, to which all the short-range
interactions are confined, and an external zone, where instead 
either no potential is present or there is a long range Coulomb or dipole
potential (or both),
 and the behavior of the solutions of the Schr\"{o}dinger equation is
very simple. In some studies, other long-range multipole potentials are
included in the external zone. \cite{{Esry_Greene:PRL99},{Seaton_Badnell:JPB99}} 
We use the $R$-matrix method in the eigenchannel form.{\cite{Greene:rev96}}
In this case we seek those stationary states for which the logarithmic
derivative of the wavefunction at the surface of
the $R$-matrix box is constant at every point. Refs.
\onlinecite{{Greene:FPADyn},{LeRouzo:84},{Fano_Lee:73}} derive a new
variational principle,

\be
{b\equiv - \frac{\partial{\log {(r\Psi_{\beta})}}}{\partial r} = 2 \frac
{\int_V {\Psi^{*}(E-\hat{H} -\hat{L}) \Psi dV}}{\int_V{\Psi^{*}
\delta(r-r_0)\Psi dV}}},
\ee
for the logarithmic derivative of the wavefunction.
If $\Psi$ is discretized in some basis set inside a spherical box, within which all the short range
dynamics is localized, this results in a generalized eigenvalue problem for
$b$:
\be
\label{eigenvalue}
\underline{\Gamma}\vec{C}=({E-\underline{H}-\underline{L}}) \vec{C} = \underline{\Lambda} \vec{C} b
\ee
where $\underline{\Lambda}$ is the  overlap of the basis functions calculated on the surface of the $R$-matrix box
and $\hat{L}$ is the Bloch operator,defined as 
\be
\hat{L}=\frac{1}{2}\delta(r-r_{0})\frac{\partial}{\partial r} r
\ee
and $r_{0}$ is the radius of the box.  The eigenvector $\vec{C}$ represents the
expansion coefficients of the basis set used. 
Both $\underline{\Gamma}$ and $\underline{\Lambda}$ are defined in the appendix
for the
finite element basis set used in this work.
It is possible to partition the basis functions in two subspaces, closed and
open, depending on whether their
value at the surface of the box is zero or nonzero.\cite{Greene:rev96} This allows us to
reduce the burden of the solution of Eq. \ref{eigenvalue}
to the easier task of solving a much smaller eigenvalue problem of type
\be
\label{eigenvalue_open} 
\Omega \vec{C}_o =
(\underline{\Gamma}_{oo}-\underline{\Gamma}_{oc}\underline{\Gamma}_{cc}^{-1}\underline{\Gamma}_{co})
\vec{C}_o = \underline{\Lambda}_{oo} \vec{C}_{o} b
\ee
in the open functions subspace,  in addition to the large auxiliary  system of
equations: 
\be
\label{linear}
\underline{\Gamma}_{cc}\vec{ C}_{c}= -\underline{\Gamma}_{co} \vec{C}_{o} 
\ee
where the subscripts indicate the matrix blocks. 
At the boundary of the $R$-matrix  box $\Psi $ is matched to an external solution
depending on the long range tail of the potential (Bessel functions for neutral molecules,  Coulomb
functions for molecular ions). This allows us to calculate
the reaction matrix $\underline{K}$, from which the scattering matrix is derived as
\be
\underline{S}=\frac{\underline{1}+i\underline{K}}{\underline{1}-i\underline{K}}
\ee
Scattering cross sections can then be calculated in the standard manner.


 \subsection{Finite element method}
The essence of the finite element method is the use of a basis set that is defined over small
local regions.
By this we mean that each basis function is nonzero only within a small region, and
it has a simple polynomial form. By using many ``sectors" or ``elements" (the
volume over which the
local basis function is defined) though, it is possible to reproduce
very complex features of the solutions to the differential equation of interest. 
We discretize $\Psi$ using finite element polynomials in all three dimensions. The
basis set is a direct product of 4 cubic Hermite polynomials defined locally in each
sector for each dimension.
The use of a spherical coordinate grid, in $r,\theta,\phi$ 
makes the boundaries of the sectors simple and the three-dimensional integrals (the main
bottleneck of these calculations) faster to calculate.


The finite element basis set is composed of  
piecewise polynomials, which  provides advantages over 
a global variable representation. In particular one can treat 
potentials and wavefunctions of complicated form by simply reducing the
size of the elements in which the polynomials are defined, in those areas where
fine features arise. In our case the basis functions are third
order Hermite polynomials, which allow us to achieve function and derivative continuity,
while still permitting a simpler implementation compared to higher order
polynomials. Each polynomial is defined in a hexahedral sector (a cube in the rescaled
variables used for the evaluation of the integrals), and since the wavefunction is
discretized in terms of finite elements in all three dimensions,
 the basis set is a direct product of 4 polynomials in each dimension
per sector, which means 64 basis functions are defined in each sector. 

In finite element analysis (FEA) the polynomials are  matched with the ones in
neighboring sectors to ensure functional and derivative continuity (and mixed
derivative continuity also, in multidimensional FEA). Each sector has 8
physical nodes (at the edges of the cube) and the basis functions defined in
the sector have coefficients (to be determined by the solution of the Schr\"{o}dinger
equation) that represent the value of the wavefunction, or its derivatives,
at the nodal point. In the language of finite element analysis, a node is  the vertex of one of the sectors
into which the three-dimensional space is divided. The matching at the boundary of
each  sector  is imposed
when assembling the global Hamiltonian matrix from the local ones.
The global index of functions that correspond to the same node and quantity (e.
g. derivative) in neighboring sectors has to be the same. Their matrix elements
have hence to be summed together.
Details of the procedure are given in the appendix.
\begin{figure}
\centerline{\includegraphics[width=12.5cm]{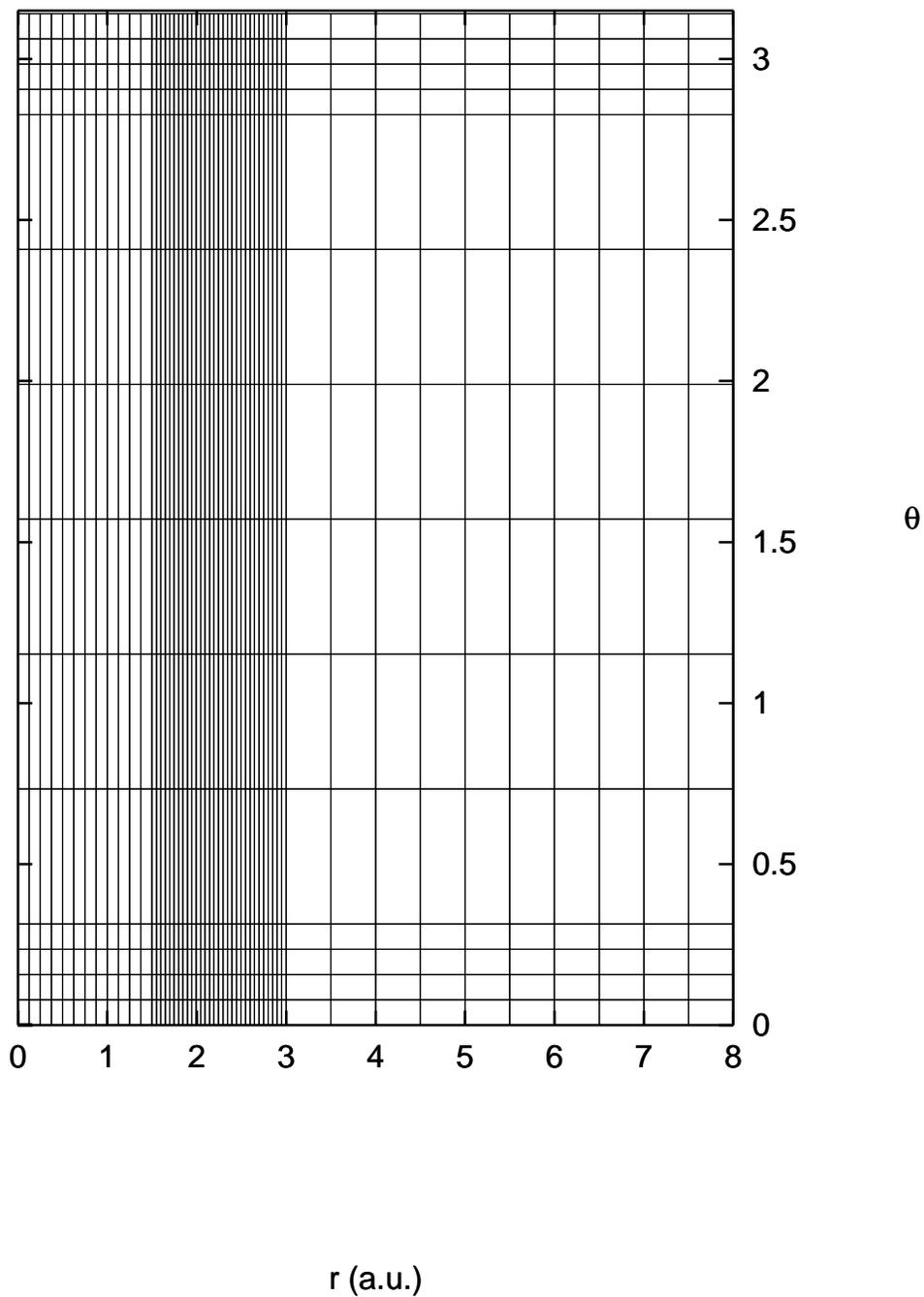}}
\caption{ From this two dimensional cut in the radius $r$ and the polar angle
$\theta$ of the finite element grid (for a $\text{CO}_{2}$
target),  it is possible to
notice the finer mesh near the oxygen nuclei localized at
$r=2.19$ a.u. and $\theta=0$ and $\pi$ respectively, while the carbon is located at
the center of the grid.} \label{grid:2D}
\end{figure}


         \subsection{ Local Density Approximation (LDA)}


Using an approach derived from Refs. \onlinecite{{Dill:2_80},{Morr_Coll:PRA78}}, we
approximate the exchange integral (that is nonlocal), by a local form  using free electron gas (FEG)
orbitals, \cite{Kohn:PR65} i.e. plane waves, for the target molecule and using the
first order Born approximation 
\be
\phi_{0} = N e^{i \vec{k}. \vec{r}}
\ee
for the scattered electron. The arbitrary normalization constant $N$ is
unimportant and it disappears as soon as we express the exchange functional as
a product of a local exchange potential times the scattered wave. After these
substitutions are made, it is possible to evaluate the integral on the
right hand side of Eq. \ref{1-body:eqn}
analytically, obtaining a local potential of the form
\be
\label{Pot:exch}
V_{ex}(\vec{r}) = -\frac{2}{\pi} k_{F} F(\eta),
\ee
whereas the Fermi momentum $k_{F}$ (the momentum of the electron that is at the
top of the Fermi sea in a free electron gas) is:
\be
k_{F}(\vec{r})=(3 \pi ^{2} \rho(\vec{r}))^{1/3}.
\ee
The other functions present in Eq. \ref{Pot:exch} are
\be
\label{Func_exch}
F(\eta)=\frac{1}{2}+\frac{1-\eta^2}{4 \eta}
\log{\left|{\frac{1+\eta}{1-\eta}} \right|}
\ee
\be
\eta=\frac{{k}}{k_{F}},
\ee
where ${k}$ is the modulus of the momentum of the scattered electron. It should
be noticed that the exchange potential in Eq. \ref{Pot:exch} is energy
dependent.

Many functionals of this form exist, \cite{Morr_Coll:PRA78} with minor
differences in
the expression for $k$, the scattering electron wavenumber.
The functional  we have used most successfully is the Hara exchange
\cite{Hara:69} where 
\be
\label{Hara:eqn}
k=\sqrt{2(E+I)+k_{F}^{2}}
\ee
and $I$ is the ionization
energy of the molecule while $E$ is the energy of the incident electron, this emerges from the assumption that the scattered
electron and the electron in the highest energy bound state (the Fermi
electron, which has momentum $k_F$)  move in the same potential field; $V_{ex}$
then depends only on $\vec{r}$, through the  electron density $\rho(\vec{r})$,
as a local potential, and on the energy, through the functional dependence of
the momentum $k$ as approximated in Eq. \ref{Hara:eqn}.

We have also experimented with 
other functional forms of the exchange interaction (still based on a FEG
approximation). One in particular is the  Slater exchange,
\cite{Slater:book} 
derived by averaging the function $F(\eta)$ over the momenta of all the
electrons up to the Fermi level, which has  often been used to calculate bound states in atoms
and molecules. However
the results using Slater exchange are unsatisfactory, presumably owing to the
neglect of the energy dependence in this model. 

Since our main goal
is to treat low energy scattering processes (0-10 eV) we linearize the
energy dependence of the functional in Eq. {\ref{Func_exch}}, in order to
calculate the exchange potential matrix
elements at all energies at once. For a molecule like $\text{CO}_{2}$, the
matrix element calculation requires around 2 hours on an
Alpha 500 Mhz workstation. The
next step is the solution of the linear system and the determination of the
scattering observables, which requires  approximately 15 minutes per energy
desired, for a basis set size of 33000.
This step is trivially parallelizable, of course.
The results improve upon inclusion of a polarization potential 
\be
\label{polar:potential}
V_{pol} = -\frac{1}{2 r^{4}}(\alpha_{0}+\alpha_{2} P_{2}(\cos{\theta}))
(1-e^{-(\frac{r}{r_{c}})^{6}})
\ee
where $r_{c}$ is a distance parameter comparable to the range of the target
charge distribution. When high accuracy is needed for resonance positions in
some applications, $r_{c}$
can be determined empirically \cite{Morr_Coll:PRA78} to
reproduce the energies of one or more resonances of interest.

All the information needed to construct the potential matrix can be extracted
from standard $ab$ $initio$ quantum chemistry codes; in this work we have used
{\sc{GAUSSIAN 98}}. The electrostatic
potential and the electronic density (needed to construct the exchange
functional) for the target molecule are calculated on a uniform cubic grid at a
CI (singles and doubles) level for the molecules
presented here. The difference in using an electrostatic potential and density
calculated at the RHF level or at the CI level for $\text{CO}_{2}$ at its
equilibrium geometry amounts roughly to a  difference of 10\% in the calculated
phase shifts and overall magnitude of the elastic cross sections. 
These calculations usually require a minimal amount of time, of the order of ten
minutes per nuclear geometry for $\text{CO}_{2}$ on the aforementioned computational platform. The potentials are
then interpolated on the three-dimensional quadrature grid using fifth order splines. 

\begin{figure}
\centerline{\includegraphics[width=20cm]{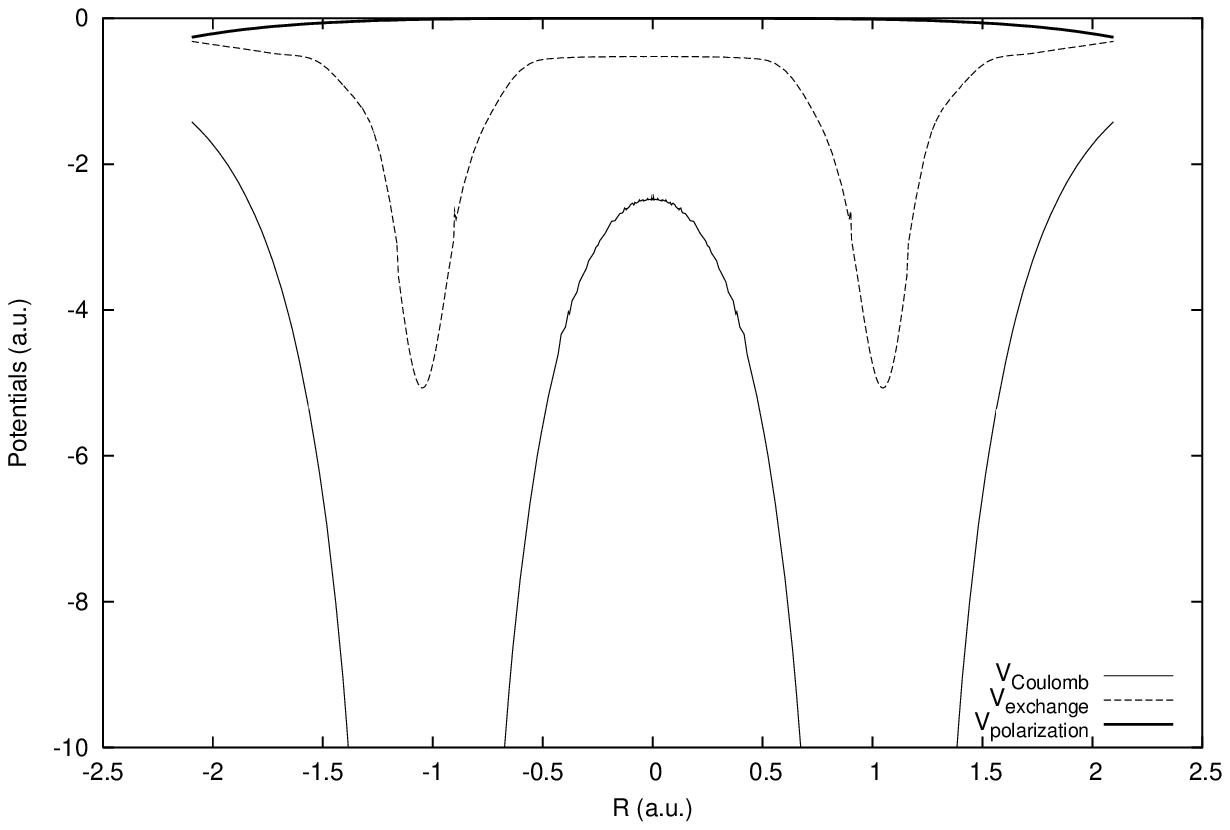}}
\caption{The three terms of the potential for a $\text{N}_{2}$ molecule. The exchange
potential is large only at the nuclei (at $r=-1.094$ and $r=1.094$ a.u. in the
equilibrium configuration of the molecule) where
the static potential is singular, so $V_{ex}$ is always much smaller than
$V_{s}$. On the other hand the polarization potential becomes important in the
outer zone, where the electron density of the molecule goes to zero.}\label{potentials:N2}
\end{figure}

\subsection {Computational Details}
The three-dimensional integrals, as was mentioned above, are the bottleneck of the
entire procedure, making it highly desirable to minimize the time spent in
their calculation.
For the sectors that do not contain a nucleus it is possible to use just
4 Gauss-Legendre points of integration, since doubling the number of points
changes the calculated phase shifts by only about $10^{-6}$ radians, 
while increasing the computational time by approximately an order of magnitude. 
Particular caution has to be observed when integrating over sectors that  contain
a nucleus. We have found it important in general to have a finite element
vertex on the Coulomb singularity, in order to obtain correct results, and to use more integration
points. In these sectors we use 20 integration points in each dimension since we found that the
convergence of the phase shifts in this case is, as in the previous case, about $10^{-6}$.

The sparse structure of the finite element matrices (see Fig.
\ref{matrix:struct}) can be exploited with great
advantage from the beginning. No matter how the grid is defined, each basis
function has matrix elements with at most 216 functions.
 This allows us to know the data structure of the matrix $\Gamma$ in Eq.
\ref{eigenvalue} in advance and store just the nonzero elements, with a
reduction of memory cost of approximately two orders of magnitude. 
This economy is crucial to allow us 
to perform three dimensional calculations in the first place.

The dimension $N$ of the eigensystem in Eq. \ref{eigenvalue} is, for
$\text{CO}_{2}$, of the order of $40000$, whereas for the open subspace it is
only $100$ or less. $N$ increases rapidly with the complexity
and spatial extension of the molecular potential, but the sparsity of the
matrices is high (about $0.5\%$ full for $N \sim 40000$ ), and  it increases with
the
dimension of the system. Depending on $N$ we use different techniques to solve
the linear system in Eq. {\ref{linear}} :
for  small $N$ we use direct sparse
LU factorization solvers (SuperLU); otherwise iterative biconjugate
gradient methods are used. Different preconditioners have been tried in this
context to speed up the solution of the linear system, the one we have found to
work the best for us is an incomplete Choleski factorization, which reduces
drastically the number of iterations with respect to a diagonal preconditioner,
the $\underline{\Gamma}$ matrix in Eq. \ref{linear} is not, in fact, diagonally
dominant. 
Clearly, the degree to which the factorization is carried out influences its
nonzero structure. The factorization is carried out to the extent that the
original structure is preserved.

Iterative methods are slower than direct
factorization, in the tests we have performed normally the direct method is
faster by a factor of ten, but for large systems an iterative solver is
essentially the only
option, owing to memory limitations. Since the factorization of a sparse matrix does not preserve the
sparsity pattern, the factorized  matrices 
present storage problems,
since a fill-in factor of
around 10 is common for these systems.

\section{ Results}


 \subsection{{Neutral molecules}}
\begin{figure}
\centerline{\includegraphics[width=12.5cm,angle=-90]{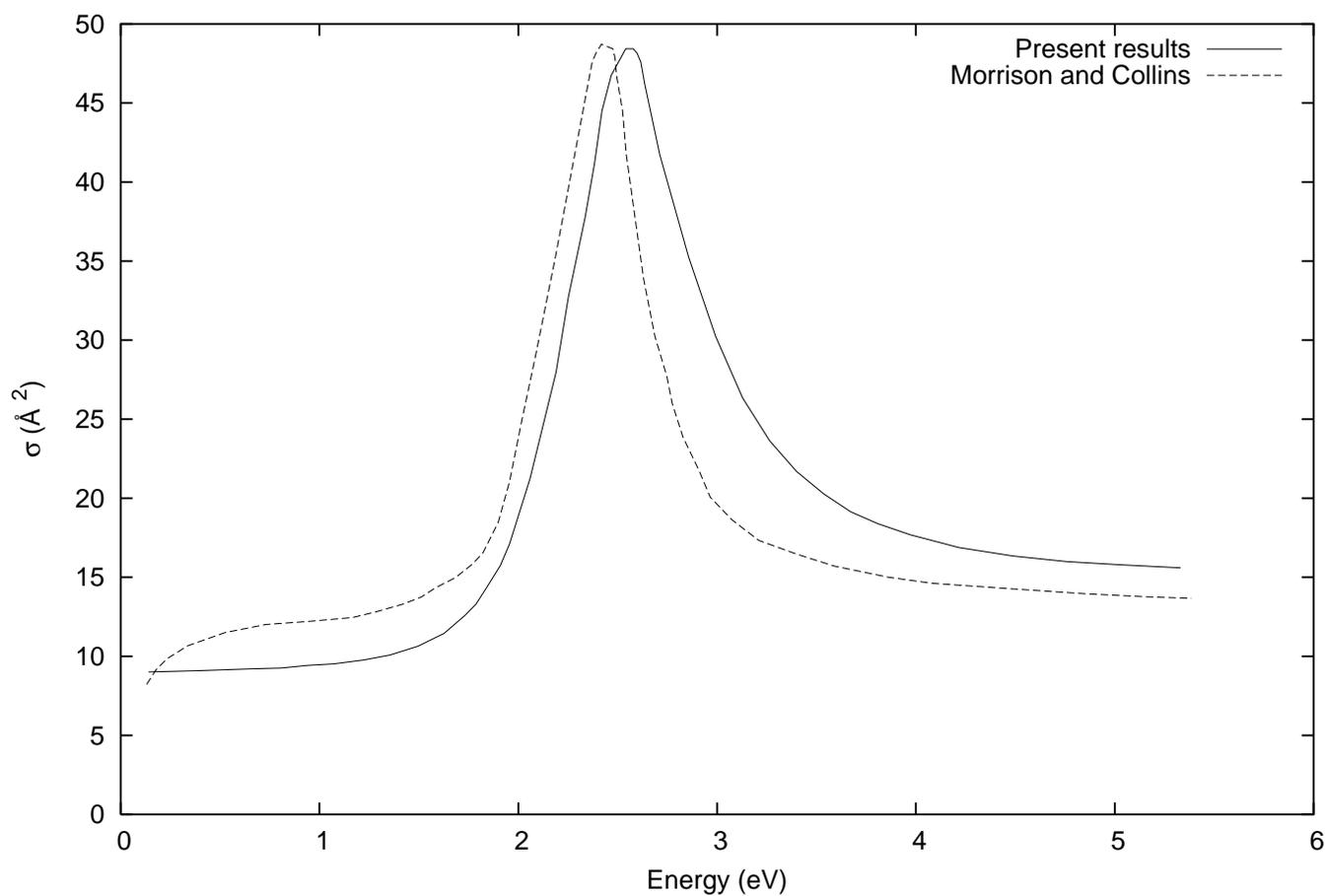}}
\caption{Total elastic cross section for electron-$\text{N}_{2}$ scattering, compared to the theoretical
results of Morrison and Collins. \cite{Morr_Coll:PRA78} } \label{n2:cross}
\end{figure}

We have tested our approach in calculations of electron scattering by
$\text{N}_{2}$ and
$\text{CO}_{2}$,  classic benchmarks in this field,
\cite{{Lynch:79},{Morr_Coll:PRA78},{Resc:99}} because their elastic cross sections
exhibit striking
features that can be challenging to reproduce. The strong and narrow $\Pi_{g}$ resonance
at 2.4 eV in $\text{N}_{2}$ is reproduced in our calculations at the right energy,
provided we use a physically reasonable  
cutoff radius $r_{c}=2.8$  a.u. for the polarization potential. The results are
shown in Fig. \ref{n2:cross}. The resonance  is reproduced also at the static exchange
level (without using a long range polarization potential), but at an energy
higher by approximately 1.5 eV. 

\begin{figure}
\centerline{\includegraphics[width=20cm]{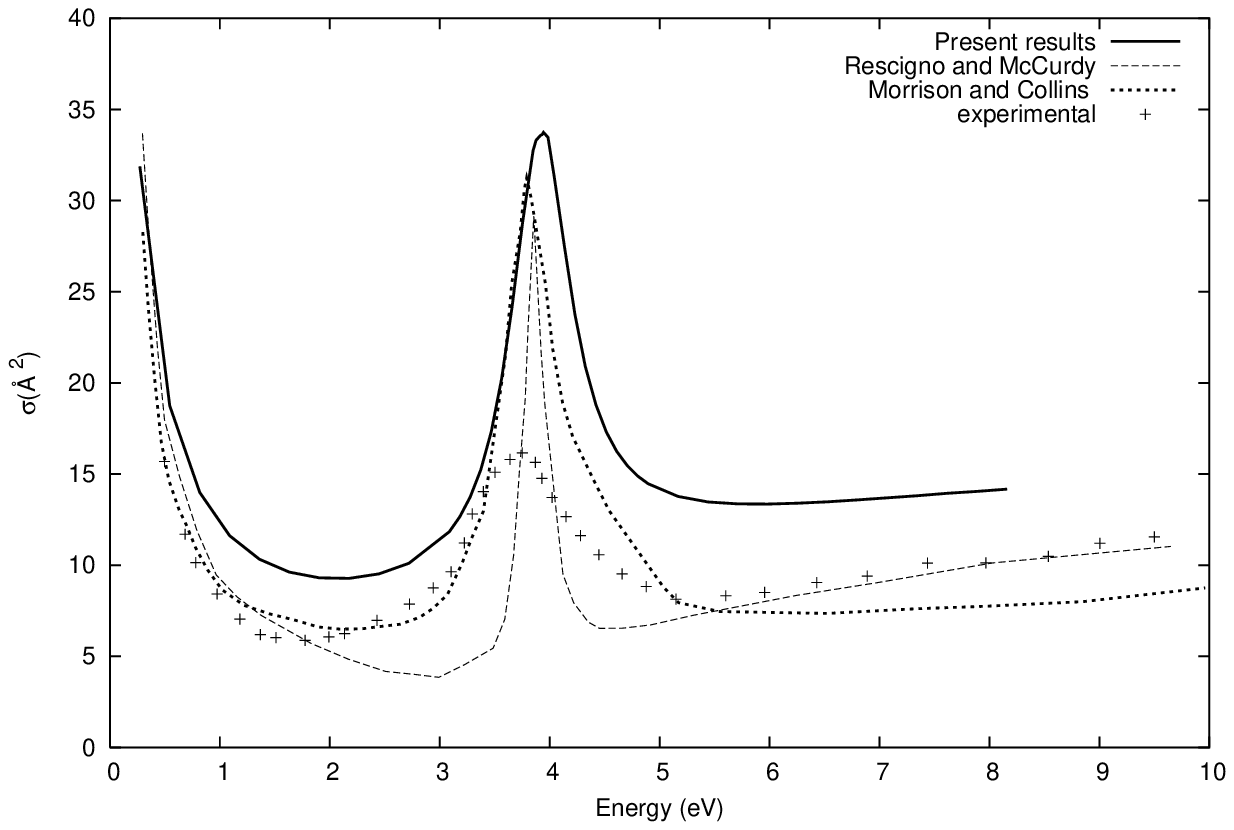}}
\caption{ Total elastic cross section for scattering of electrons from
$\text{CO}_2$. The
present results are compared with previous theory from Rescigno $et \;al.$
\cite{Resc:99} and Morrison and Lane, \cite{Lane:rev80} whereas the
experimental results are those of Szmytkowski. \cite{Szmitkowski:co2}}
\label{co2:cross1}
\end{figure}

For $\text{CO}_{2}$ the main feature in the total
elastic cross section is a $\Pi_{u}$ resonance at 3.8 eV. To reproduce it at the
correct energy we have to tune the polarization cutoff radius to $2.4$  a.u.. 
This feature is present also at the static exchange level, at 8 eV. The
dependence on the polarization, as one expects from the larger spatial
extension, the larger number of electrons and  the greater asymmetry of this
molecule, becomes much more
pronounced than in $\text{N}_{2}$. The scattering cross section for this system is shown in Fig.
\ref{co2:cross1} . The value of the cutoff radius for the polarizability
potential, which is the only adjustable parameter in the model, is
reasonable. This is clear from Fig. \ref{potentials:N2} which demonstrates that this
potential is appreciable just outside the region
where the main part of the electronic density is located. 
The results are always in
good agreement with previous theory, as shown in the figures. Vibrational
effects tend to broaden these resonances in experimental elastic scattering cross
sections, and they also give rise to more structured resonance peaks, which are
not considered in this work.
The present calculations have been performed for the molecular targets only at
their equilibrium distances.
The values of the polarizabilities used in these calculations
are \cite{Morr_Coll:PRA78}
$\alpha_{0}=11.89a_{0}^{3}$ and $\alpha_{2}=4.19a_{0}^{3}$ for $\text{N}_{2}$
and \cite{Lane:rev80} $\alpha_{0}=17.9a_{0}^{3}$,
$\alpha_{2}=9.19a_{0}^{3}$ for $\text{CO}_{2}$. It should be pointed out that
accurate static polarizability coefficients $\alpha_{0}$ and $\alpha_{2}$ in Eq.
\ref{polar:potential} can also be
extracted from  $ab$ $initio$ calculations. 
Generally, the low-lying shape resonances present in these small molecules are
spatially highly localized,  which allows the radius of the $R$-matrix box to be kept small,
around 8 to 14 a.u.
for the present calculations.

In the case of the third neutral molecule that we present here, ethylene, the situation
is more complicated. Since the target is now nonlinear it is more difficult to
describe it in a discrete basis set and it is more expensive computationally to calculate the
scattering cross section. Nevertheless we are able to reproduce the features of
the elastic cross section for this molecule. We find  good agreement with the
energies  of
the resonances and with the overall cross section magnitude, compared with previous theory and also experimental
data, although the vibrational effects again tend to broaden the resonance
peak.

It should be noticed parenthetically that if we neglect exchange altogether in
calculations for all of the
molecules presented here,
the cross sections are qualitatively wrong, with
resonances far lower in energy than the experimental ones and in the wrong symmetry
channels. This is due to the fact that some of the target electrons are not
bound anymore, because the static potential is not attractive enough. Once
added, the exchange potential 
is basically an attractive local potential, resulting in the
correct number of bound states for the target; consequently the
scattering resonances are generated by capture of the electron in truly
unoccupied molecular orbitals  of the target.
A more systematic study of the behavior of the cross sections, when
different parts of the potential are neglected altogether, can be found in Ref.
\onlinecite{Lane:rev80}. 

\begin{figure}
\centerline{\includegraphics[width=20cm]{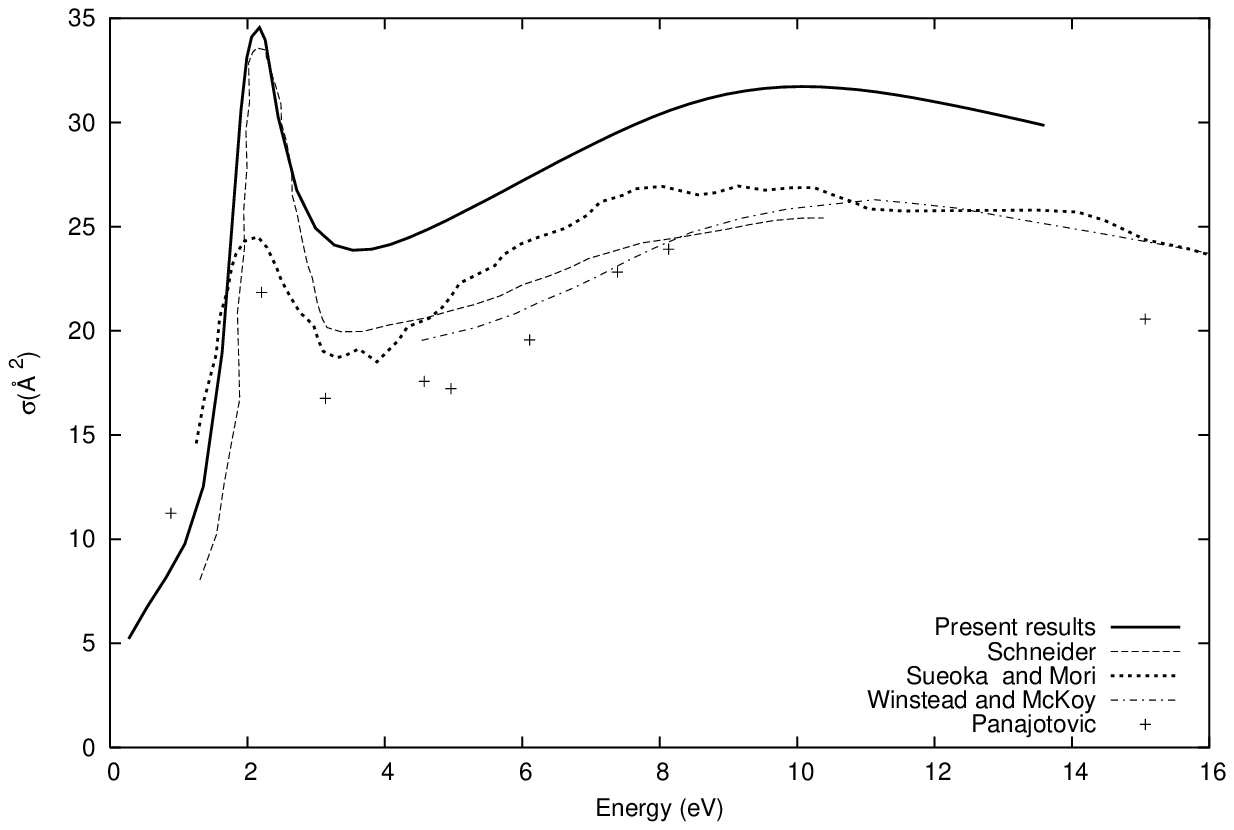}}
\caption{Total elastic cross section for electron-$\text{C}_{2}\text{H}_{4}$ scattering,
compared to previous theoretical results of Winstead $et.\;al.$ \cite{McKoy:c2h4} and of
Schneider $et.\;al.$. \cite{Schneider:c2h4} The experimental results are the ones of Panajotovic
$et.\;al.$ \cite{Panajotovic:c2h4} and of Sueoka and Mori. \cite{Sueoka_Mori:86}}
\label{c2h4:fig}
\end{figure}


 \subsection{{Quantum defect calculations}}
It has been shown {\cite{Tashiro:02}} that use of a local density approximation
can often be
effective in calculating molecular quantum defects, for bound or scattering
states, for small closed-shell target molecules. 
It is possible to calculate quantum defects from a scattering calculation
carried out near zero energy. The key step is to 
diagonalize the $K$-matrix
\bd
K_{ii'}=\sum_{\alpha}{U_{i \alpha} \tan{\pi \mu_{\alpha}} U^{T}_{\alpha i'}}
\ed
and then utilize the relationship between
the quantum defect and the  scattering phase shift,
\cite{{Seaton:Rpp83},{Greene:rev96}}
\be
\delta_{l}=\pi \mu_{l}
\ee

Accordingly quantum defects 
can be extracted from electron-scattering calculations at positive
or negative energies.
These  quantum defects can then be used to determine the Born-Oppenheimer
potential curves of the Rydberg states converging to the various ionization
thresholds through the Rydberg formula. \cite{Greene:rev96} these can then be
exploited through MQDT
techniques, to extract dynamical information on, for example, dissociative
recombination, \cite{Slava:03} a process that we will study  in the future using the machinery developed in this
paper.
Here we show an example of how well this approach works for a simple diatomic
molecule. 

\begin{figure}
\centerline{\includegraphics[width=20cm]{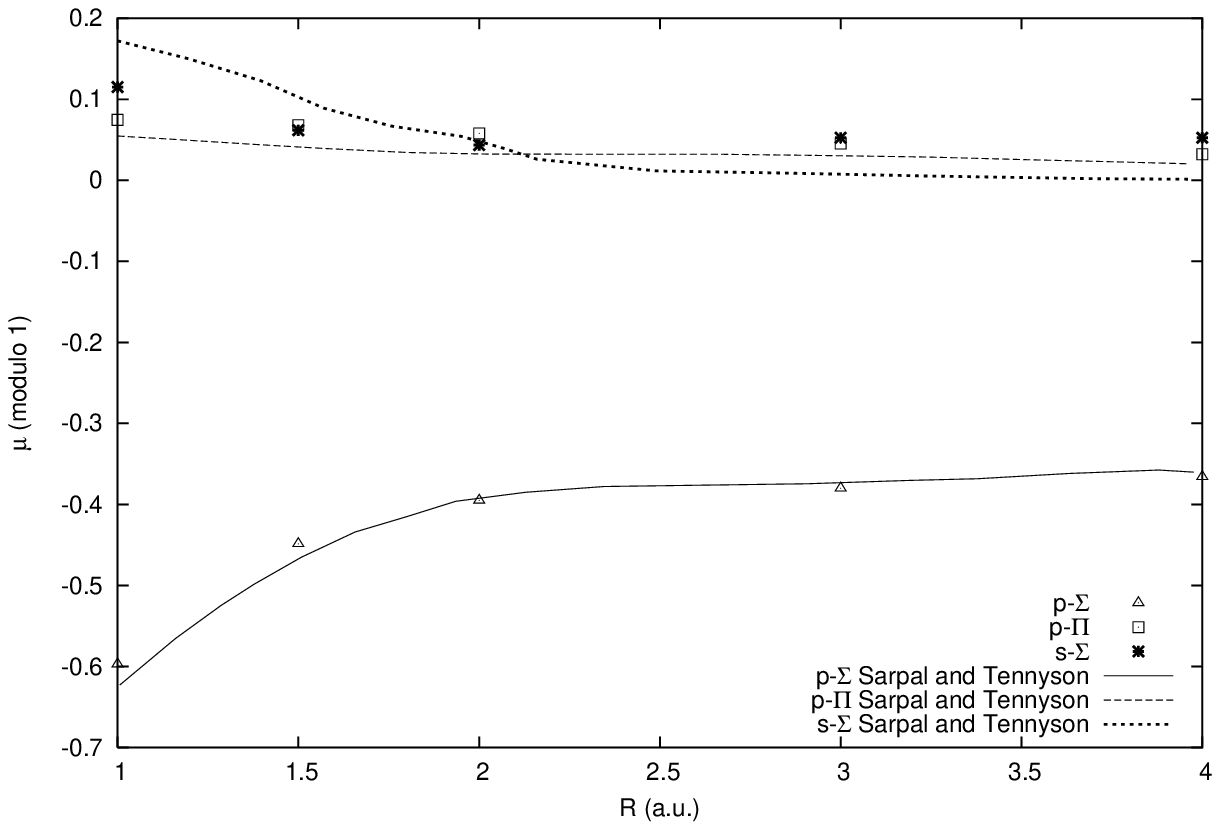}}
\caption{Comparison of quantum defects for the  $\text{HeH}$ molecule
calculated with our method to the calculations of Sarpal and Tennyson.\cite{Tenn:92}}\label{Tenn:qd}
\end{figure}

We compare our results to the work of Sarpal and Tennyson 
\cite{Tenn:92} which made
no approximation about the nature of the electron-molecule potential. It is possible to see that
the agreement is generally very good. The quantum defects represented in Fig.
\ref{Tenn:qd}
are the most important ones, higher symmetries and partial waves ($l>2$) having very 
small phase shifts at the low energies considered here. In electron scattering
from an ionic target we must 
account for the fact that heteronuclear molecules  like $\text{HeH}$ have a dipole moment, so
we must transform from the center of mass frame to a new frame centered on
the center of charge (the proton in this case). It is then possible 
to match to simple Coulomb functions at the boundary of the box. Otherwise
multipole potentials have to be included in the external region.


 \section{  Conclusions}
In this paper we have shown how a combination of the $R$-matrix method and
a three-dimensional finite element basis set can provide a promising tool for solving problems in
which a low-energy
electron collides with a polyatomic molecule. It should be emphasized that to
perform three-dimensional calculations in a local basis set there is need for
special computational techniques, namely sparse matrix techniques.
These calculations are in
general very complicated and time consuming, so some approximation must be
made in order to make them sufficiently manageable. In the present work we
approximate
the exchange term in the potential, which is nonlocal, as a local potential
using the free electron gas approximation. The results are shown to be
qualitatively accurate for a number of molecules even in this rather crude
approximation. Nevertheless there is room for improvement for further work directed at
treating exchange exactly and including relaxation of the target
orbitals in the presence 
of the scattering electron.

\section*{Acknowledgments}
This work was supported by the Department of Energy, Office of Science, and by
an allocation of NERSC supercomputing resources. We
thank J. Shertzer for useful discussions at an early stage of
the project. We have also benefited from a number of useful discussions with
R. Santra.

\appendix
\section{Finite Element matrices}
\label{App:A}

\begin{figure}
\centerline{\includegraphics[width=12.5cm]{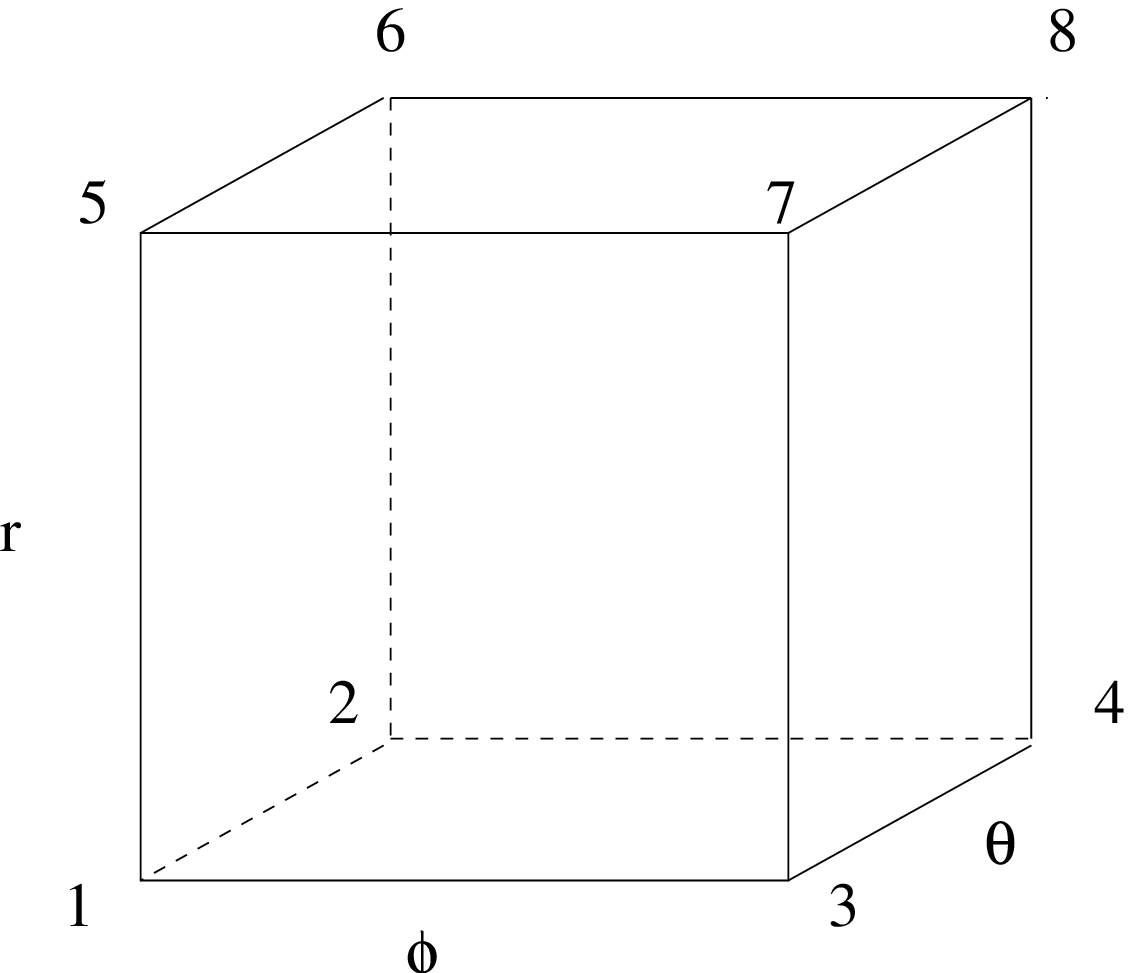}}
\caption{Nodal structure for each finite element sector: indicated are the
spherical coordinates and the numbering of the nodes at the vertices of the
sector.}\label{cube:nodes}
\end{figure}

Starting from Eq. \ref{eigenvalue} we define the matrices $\Gamma$ and $\Lambda$ in our finite element
basis transforming first to spherical coordinates (the box is spherical and the
grid is also defined in spherical coordinates), and then to rescaled
coordinates, which are
the variables of the local polynomials. In the rescaled variables each sector
is transformed to a cube, in which the range of each variable is from $0$ to
$1$. 
The nodal structure of each element is
represented in Fig. \ref{cube:nodes} and the wavefunction inside each sector can
be expanded as 
\be
u(\xi_{1},\xi_{2},\xi_{3}) = \sum_{i,j,k,l,m,n} \psi_{i}^{l}(\xi_{1})
\psi_{j}^{m}(\xi_2) \psi_{k}^{n}(\xi_3) C^{(lmn)}_{node}
\ee
where $i,\;j,\;k$ can be $1$ if the polynomial has nonzero value at some node or
$2$ if it has nonzero derivative, whereas $l,\;m,\;n$ can assume values of $0$ if
that node is the first for the variable of the polynomial in the sector or $1$
if it is the last; $\xi_{i}$  are the local rescaled variables .
The coefficients $C^{(lmn)}_{node}$ are the values of the wavefunction and its
derivatives at the node, and they are to be determined solving Eq. \ref{eigenvalue}.
If we define 
\be
a_{k,p}=x_{k,p,i+1}-x_{k,p,i}
\ee
\be
x_{k,p}=a_{k,p}\:\xi_{k}+x_{k,p,i}
\ee
where $k$ indexes the spherical coordinates and $p$ the sectors in which they
are defined, $x_{k,p,i}$ and $ x_{k,p,i+1}$ are the initial and final points
for the variable $x_{k}$ in sector $p$, the expressions for the matrices
become:
\be
\label{Gamma:app}
\Gamma_{ij}=\int{\left[\sum_{k}^{3} \frac{F(x_k)}{a_{k}a_{k}} \frac{\partial
u_{i}}{\partial
\xi_{k}} \frac{\partial u_{j}}{\partial \xi_{k}} + 2 u_{i}(U-E)u_{j} \right]  a_{r} a_{\theta} a_{\phi} r^{2}\sin^{2}{\theta} d\xi_{1} d\xi_{2} d\xi_{3}   }
\ee
\be
\label{Lambda:app}
\Lambda_{mn} = \int{Y^{*}_{lm}(\theta, \phi) Y_{l'm'}(\theta, \phi)
\sin{\theta}\;d \theta\; d \phi} =
\delta_{ll'}\delta_{mm'}
\ee
where $F(x_k)$ is a spherical coordinates scale factor, and it is $1$ if $x_{k}=r$ and $1/r^2$ and $1/(r^2 \sin^2{\theta})$ for
$\theta$ and $\phi$ respectively.
Imposing function and derivative continuity for $u(\xi_{1},\xi_{2},\xi_{3})$
amounts to require that the indices of the same node across neighboring sectors
be the same. This in turn leads to having to perform a sum of the integrals in
Eq. \ref{Gamma:app} when evaluating the matrix element at a node, across all
sectors that share that node.

\begin{figure}
\centerline{\includegraphics[width=12.5cm,angle=-90]{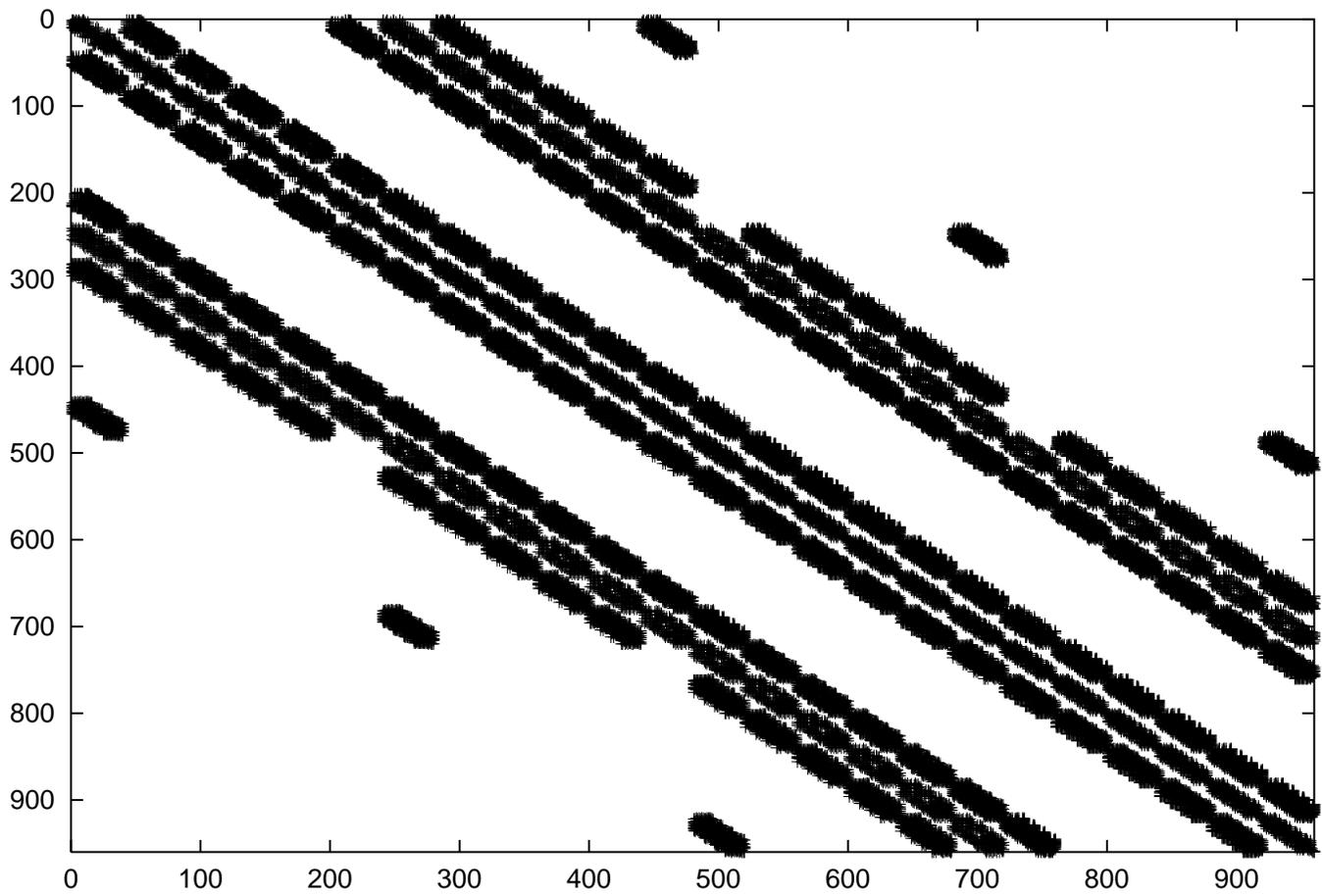}}
\caption{Structure of the finite element matrix $\Gamma$ for a small test case of
dimension 900. It is possible to notice the great sparsity of the matrix, which
increases with the dimension of the matrix.}
\label{matrix:struct}
\end{figure}

\bibliography{paper_FEA}

\printfigures
\end{document}